\documentclass[%
reprint,
showpacs,preprintnumbers,
 amsmath,amssymb,
 aps,
 prl,
]{revtex4-1}

\usepackage{epsfig}
\usepackage{psfrag}
\usepackage{verbatim}
\usepackage[usenames]{color}

\newcommand{\pullback}[1]{\hspace{-#1pt}}
\newcommand{\be}{\begin{equation}}
\newcommand{\ee}{\end{equation}}
\newcommand{\lc}{{\ell_\mathrm{c}}}
\newcommand{\lp}{{\ell_\mathrm{p}}}
\newcommand{\kBT}{{k_{\textrm{B}} T}}
\newcommand{\rr}{\mathbf{r}}

\newcommand{\ww}{\mathbf{w}}
\newcommand{\pp}{\mathbf{p}}

\newcommand{\RR}{\mathbf{R}}
\newcommand{\JJ}{\mathbf{J}}
\newcommand{\mbg}{\mathbf{g}}
\newcommand{\ns}{n_\textrm{s}}
\newcommand{\energy}{\mathcal{E}}
\newcommand{\IDM}{\mathbf{1}_n}
\newcommand{\ttt}{\mathbf{t}}
\newcommand{\DDelta}{\mathbf{\Delta}}
\newcommand{\ddelta}{\boldsymbol{\delta}}
\newcommand{\gd}[1]{\mathbf{g}_{#1}}
\newcommand{\Gd}[1]{\mathbf{G}_{#1}}

\newcommand{\ddxn}[3]{\frac{\partial^#3 #1}{\partial #2^#3}}
\newcommand{\dx}[1]{\mathrm{d} #1}
\newcommand{\del}[1]{\partial #1}
\newcommand{\avg}[1]{\langle #1 \rangle}
\def\lsim{\mathrel{\rlap{\lower4pt\hbox{$\sim$}}
    \raise1pt\hbox{$<$}}}                % less than or approx. symbol
\def\gsim{\mathrel{\rlap{\lower4pt\hbox{$\sim$}}
    \raise1pt\hbox{$>$}}}                % greater than or approx. symbol

\begin{document}

\title{Off-lattice Monte Carlo Simulation of Supramolecular Polymer Architectures}
\author{H.E. Amuasi$^{1}$, C. Storm$^{1,2}$\\
 \normalsize{$^1$Department of Applied Physics and $^2$Institute for Complex
    Molecular Systems,\\Eindhoven University of
    Technology, P. O. Box 513, NL-5600 MB Eindhoven, The Netherlands}\\}

\begin{abstract}
We introduce an efficient, scalable Monte Carlo algorithm to simulate cross-linked architectures of freely-jointed and discrete worm-like chains.
Bond movement is based on the discrete tractrix construction, which effects conformational changes that {\em exactly} preserve fixed-length constraints of all bonds. The algorithm reproduces known end-to-end distance distributions for simple, analytically tractable systems of cross-linked stiff and freely jointed polymers flawlessly, and is used to determine the effective persistence length of short bundles of semi-flexible worm-like chains, cross-linked to each other. It reveals a possible regulatory mechanism in bundled networks: the effective persistence of bundles is controlled by the linker density.
\end{abstract}


\maketitle

The Kratky-Porod worm-like chain (WLC) \cite{KP1949} has proven to be an indispensable model for the coarse-grained description of stiff polymers.  Biophysicists, in particular, have applied the model to glean the mechanics of a large variety of biological filaments, including actin \cite{MacKintosh,Bausch}, double-stranded DNA \cite{MarkoSiggia95,BustaScience}, unstructured RNA \cite{Liphardt}, fibrin \cite{Storm}, tropocollagen \cite{Sun}, and many other polypeptides.  However, in biologically relevant settings such filaments hardly ever occur or function alone as single chains.  Instead, supramolecular associations --- covalent or transient --- locking many chains into bundles or networks are the prevalent motif.  While it is perhaps only natural to consider supramolecular structures of WLC's and study the effect of cross-linking in them, attempts to do so are severely hampered by the practical incompatibility of the connectivity constraints inherent to the architecture with adaptable and fast numerical algorithms. This Letter outlines an effective, scalable method for numerical analysis of such architectures, and clears the way for precise statistical-mechanical analysis of realistic supramolecular polymeric assemblies.

The WLC is specified by a continuously differentiable space curve $\rr(s)$ of length $\lc$ parametrized by the arc length parameter $s$.  It is further endowed with a Hamiltonian that quantifies the cost of bending the curve:
\be
\label{eqn:WLCHamiltonian} \mathcal{H} = \frac{\kappa}{2}\int_0^\lc \pullback{3}\dx{s}\left( \ddxn{\rr}{s}{2}\right)^2,
\ee
where $\kappa$ is the bending modulus.  Implicit in this definition is the constraint of local inextensibility, that is, the local tangent magnitude $\left|\del{\rr}/\del{s}\right|$ is unity.  The persistence length $\lp = \kappa / \kBT$ is the characteristic length governing the decay of tangent-tangent correlations and provides a quantitative measure for a polymer's flexibility.
Though the specification of the WLC model appears to be simple, the constraint of local inextensibility inherent in the model leads to considerable mathematical difficulty when attempting to obtain an analytical solution of even the simplest of such thermally fluctuating network structures.  Nonetheless, in the so-called {\em semiflexible} ($\lp \gtrsim \lc$) limit the radial distribution function may be obtained analytically \cite{Wilhelm1996}.  Furthermore, isotropic random networks of WLC's serve as a model for the mechanics of general filamentous biomaterials \cite{MacKintosh, Storm}.

\begin{figure}
\centerline{\psfig{file = 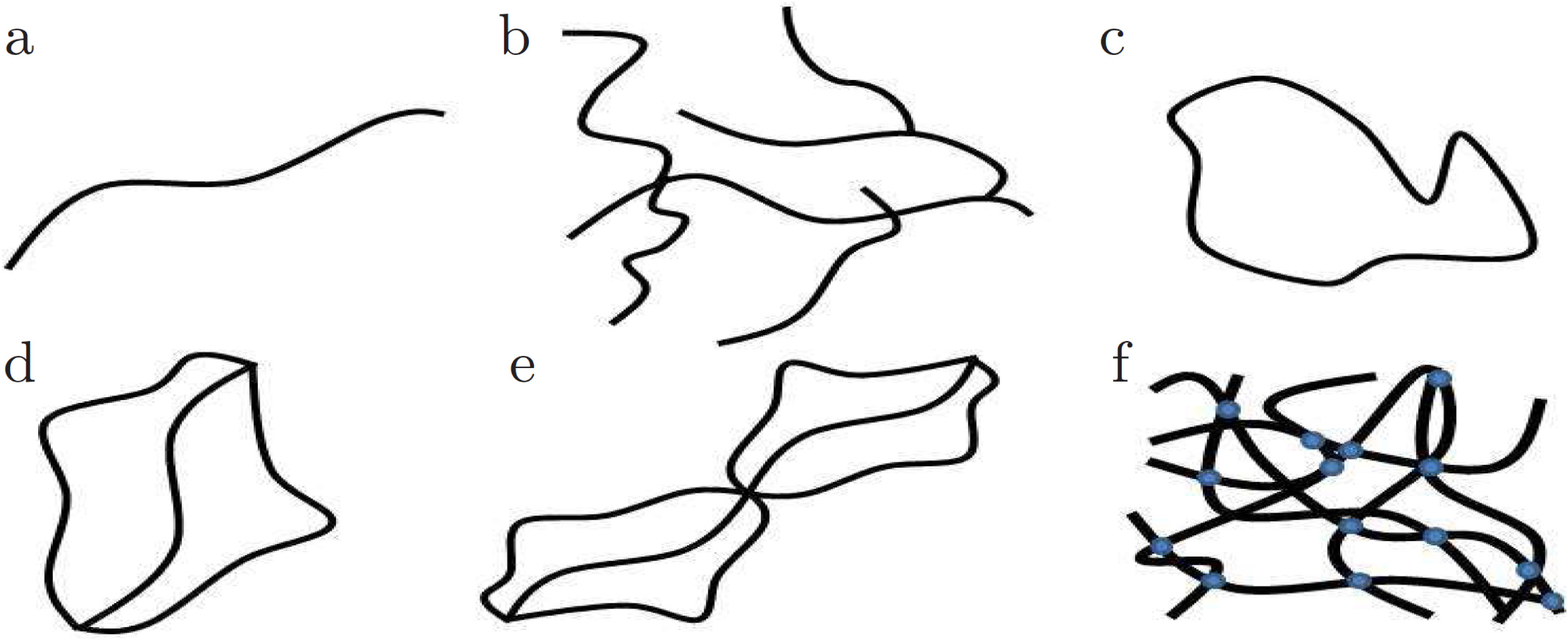, width = 1\linewidth}}
\caption{(a) Linear chain; (b) loopless branched polymer; (c) closed-loop; (d), (e) and (f) are network structures which possess two or more closed-loops that share the same polymeric link.  The MCMC move TRACTRIX, introduced in this Letter, can address these structures. }
\label{fig:1}
\end{figure}

Laboratory experiments and computer simulations may have to pave the way to investigate the properties of those biological networks that remain analytically intractable.  Even so, non-trivial complications arise since one often, as a first step, needs to discretize the WLC reducing it to a chain of tethers of fixed length (see, {\em e.g.}, \cite{StormNelsonDPC}).  The widely used Molecular Dynamics (MD) constraint algorithms, such as SHAKE \cite{shake} and RATTLE \cite{rattle} have been developed to deal with these fixed-length constraints, but have, in practice, been limited to tree-structures and rigid loops --- see Fig.~\ref{fig:1} b) and c).

Markov Chain Monte Carlo (MCMC) constraint algorithms offer a tantalizing alternative in that being purely stochastic, they allow for unphysical moves thus eliminating the need for time-step integration and quickly providing good equilibrium statistics.
Over the past 30 years, a number of `smart' MCMC moves have been advanced for the simulation of atomistic models of melts of polymeric systems \cite{frenkel2001, Dodd1993, Mavrantzas2005:HMM}.  After a few modifications (for example, after removing the  fixed bond-angle constraint inherent in many atomic-scale models) most of these techniques can be carried over into the simulation of the more coarse-grained discrete WLC models.  However, these techniques are limited in the variety of cross-linked architectures which they are able to address \cite{Mavrantzas2005:HMM, Leontidis1994}.

In this Letter we introduce TRACTRIX: a MCMC move that may be used to accurately simulate various cross-linked freely-jointed and discrete WLC architectures (Fig.~\ref{fig:1}).  The common feature of the structures (d), (e) and (f) in Fig.~\ref{fig:1} that distinguishes them from the rest [(a), (b) and (c)] is the existence of conjoined closed loops that share one or more polymer links, a feature we wish to address, and which is a central characteristic of both supramolecular filaments and cross-linked networks.

Three main technical problems are to be addressed in the implementation of our method:  ({\em i}) preservation of the connectivity of the network structure, ({\em ii}) conformance to the fixed-length constraints of the inter-linking tethers, and ({\em iii}) {\it detailed balance}.  For the latter requirement, we adhere to the recipe of the standard Metropolis algorithm, which is to ensure that each trial move is reversible, and that its probability of acceptance --- the so-called {\it acceptance ratio} --- is correctly computed to eventually yield Boltzmann statistics.  As pointed out, for instance, by Maggs \cite{Maggs2006:PRL}, in continuum systems such as ours the volume element in the vicinity of the state is also transformed, and we must therefore consider the Jacobian determinant of the transformation when determining the acceptance ratio.

\begin{figure}
\centerline{\psfig{file = 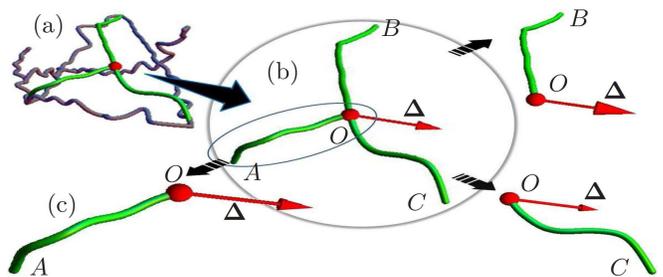, width = 1\linewidth}}
\caption{(a) Section of a network of WLC's.  (b) The basic unit of a network: the star.  When the central node $O$ is displaced by a random vector $\DDelta$, each of the arms [e.g., link $\widetilde{AO}$ in (c)] of the star may be treated independently.}
\label{fig:2}
\end{figure}

%\section{\label{sec:Intro} The Tractrix}
The basic unit of any network is the {\it star} which consists of $\ns$ linear chains terminating at a single central node (see Fig.~\ref{fig:2}) by means of a cross-link.  Any positive integer value for $\ns$ constitutes a star, but typically for biological networks $\ns = 4$.  Without loss of generality, let us consider the 3-arm star illustrated in Fig.~\ref{fig:2}b and assume that the ends $A$, $B$, and $C$ of its arms are temporarily fixed in space but that the central node $O$ is free to move.
To preserve the network connectivity at all times we must, whenever $O$ is displaced by some vector $\DDelta$, move all the ends of the linear chains that terminate at $O$ by the same displacement.  Typically $\DDelta$ is a random displacement chosen from a spherically symmetric distribution during the simulation.
We may thus treat each of the linear chains independently so our focus may now narrow down to a single linear chain anchored at one end but with the other end free to move.
The problem at hand may be set forth in two parts: firstly, given the initial contour of the chain, how may we reversibly deform it so that its free end is displaced by exactly $\DDelta$? \cite{note1} In effect, we seek a suitable invertible transform $\Gd{\DDelta}$ that will act on the chain incrementing its end-to-end vector by $\DDelta$. Secondly, with what probability must we accept this deformation?

We will proceed by considering a linear chain for which {\it both} ends are initially free to move and we will determine its new position after we displace one end by $\ddelta_0$.  The chain may be discretized  into a series of $N$ bonds $\ttt_1$, $\ttt_2$, $\ldots$, $\ttt_N$ of fixed length $t$ interlinking the sequence of coordinates $\rr_0$, $\rr_1$, $\ldots$, $\rr_N$, so that the $i$-th bond $\ttt_i = \rr_i - \rr_{i - 1}$, and
$ \left| \rr_i - \rr_{i - 1} \right| - t = 0$ ($t = \mbox{const.}$ and $i = 0,\,1,\,\ldots,\,N$).
The bending energy of the discretized WLC is given by \cite{Wilhelm1996}$\energy(\{\rr_k\}) = -\varepsilon \sum_{i = 1}^{N - 1} \ttt_i \cdot \ttt_{i + 1}$,
where $\varepsilon = \kBT \lp/t^3$.  This expression can be shown to approach the energy for the continuous chain in Eq.~\eqref{eqn:WLCHamiltonian} in the limit of $N \rightarrow \infty $ and $t \rightarrow 0$ while keeping constant $Nt = \lc$ and $\varepsilon t^2 / N$.  However, it is sufficient to discretize the WLC so that there are at least 3 bonds in one persistence length \cite{Wilhelm1996}.
Let $\gd{\ddelta_{i - 1}}$ be an operator that will transform the pair $\left\{ \rr_{i - 1},\,\rr_i\right\}$ into $\left\{ \rr'_{i - 1},\,\rr'_i\right\}$ so that $\rr'_{i - 1} = \rr_{i - 1} + \ddelta_{i - 1}$ and $\left| \rr'_{i - 1} - \rr'_i\right| = \left| \rr_{i - 1} - \rr_i\right| = t$.
One may readily select any known transformation that satisfies these two latter requirements, but for reasons that we will justify shortly, let us adopt Hoffman's \cite{Hoffman2008:DDG} {\it discrete tractrix} construction \cite{note2} which is illustrated geometrically in Fig.~\ref{fig:3}a:
first, rigidly translate the bond $\ttt_i$ by $\ddelta_{i - 1}$ so that {$\left\{ \rr_{i - 1},\,\rr_i\right\}$ is moved to $\left\{ \rr'_{i - 1},\,\rr''_i\right\}$.
Next, form the parallelogram spanned by $\ddelta_{i - 1}$ and $f\ttt_i$, where $f$ is some adjustable factor ($f = 2$ in Fig.~\ref{fig:3}).
Finally, to obtain $\rr'_i$ reflect $\rr''_i$ in the parallelogram's diagonal which passes through $\rr'_{i - 1}$.
Fig.~\ref{fig:3}b shows that following the same procedure for the inverse transform $\gd{{\ddelta}_{i - 1}}^{-1}=\gd{{\textrm{--}\ddelta}_{i - 1}}$, but this time starting from the initial position $\left\{ \rr'_{i - 1},\,\rr'_i\right\}$ we recover the original pair $\left\{ \rr_{i - 1},\,\rr_i\right\}$ thus demonstrating the transform's reversibility.
It can be shown that
\be
\label{eqn:BondTransform} \rr'_i = \mbg_{{}_{\textrm{T}}}\pullback{3}\left(\rr_i,\,\rr_{i - 1},\,\ddelta_{i - 1},\,f \right) \equiv \pp_i + 2 \ww_i \frac{\ttt_i \cdot \ww_i}{\ww_i \cdot \ww_i}
\ee
where $\pp_i = \rr'_{i - 1} - \ttt_i$ and $\ww_i = f\rr_i + (1 - f)\rr_{i - 1} - \rr'_{i - 1}$.

\begin{figure}
\centerline{\psfig{file = 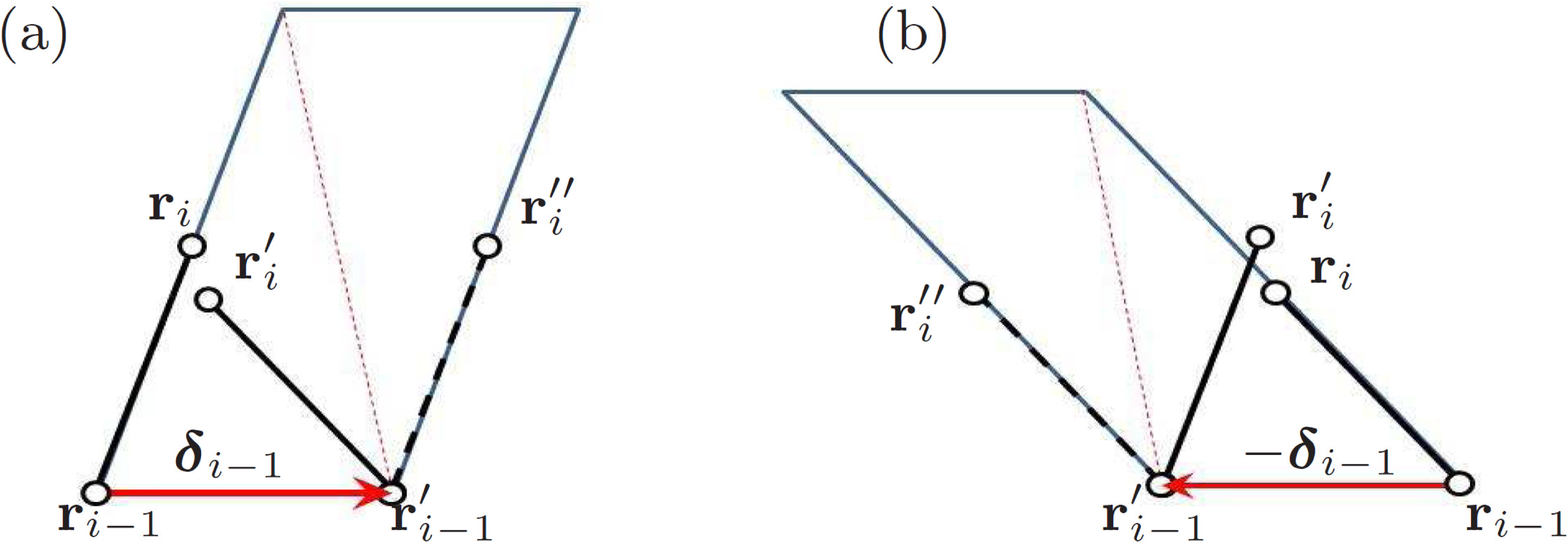,width = 1\linewidth}}
\caption{(a) Discrete tractrix transformation of the bond $\ttt_i$:
1. Rigidly translate the bond $\ttt_i$ by $\ddelta_{i - 1}$ so that $\left\{ \rr_{i - 1},\,\rr_i\right\}$ is moved to $\left\{ \rr'_{i - 1},\,\rr''_i\right\}$.
2. Construct the parallelogram spanned by $\ddelta_{i - 1}$ and $2\ttt_i$.
3. Reflect $\rr''_i$ in the parallelogram's diagonal (dashed line) to obtain $\rr'_i$.
(b) Inverse of the transformation in (a) showing that the tractrix transformation is reversible. In both cases the length of the bond is preserved.}
\label{fig:3}
\end{figure}

\begin{figure*}
\centerline{\psfig{file = 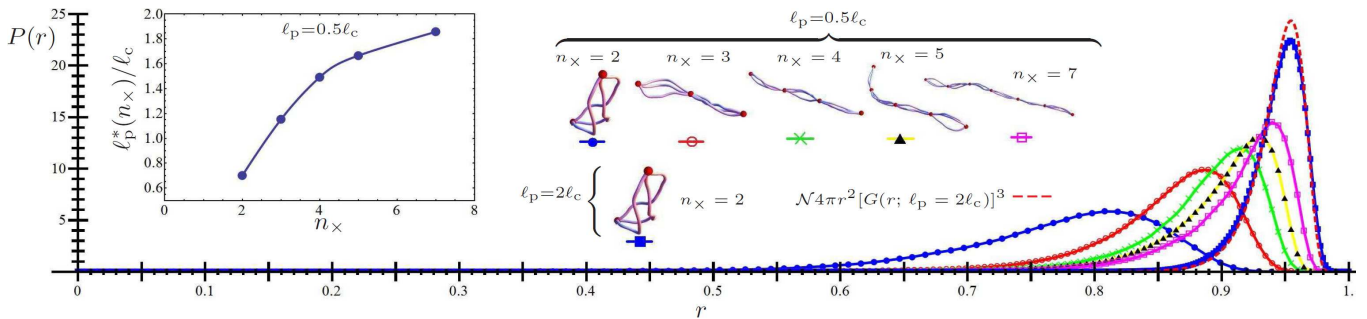, width = 1\linewidth, height=0.23\linewidth}}% Here is how to import EPS art
\vspace{-.4cm}
\caption{\label{figPDF:wide} Simulation results (after $\sim 10^7$ Monte-Carlo steps): probability density distribution functions $P(r)$ for the end-to-end distance $r$ (normalized to unity) of freely fluctuating bundles of three WLC's each with $\lp = 0.5\lc$ except for the last bundle which has $\lp = 2\lc$.  Each bundle is cross-linked at equal intervals along the contour lengths of its chains. $n_\times$ is the number of cross-links in a bundle.  The red dashed curve is an analytical estimate for the last bundle: $\mathcal{N} 4\pi r^2 [G(r)]^3$ where $G(r)$ is the spherically symmetric radial distribution function for a stiff WLC \cite{Wilhelm1996} (here $\lp =2\lc$), and $\mathcal{N}$ is a normalization constant.  As expected, this prediction agrees fairly well with results for bundles with $\lp \gsim \lc$.  The inset shows the dependence of {\it effective persistence length} $\lp^{\pullback{4}*}$ (of those bundles with $\lp = 0.5 \lc$) on $n_\times$. (See text for details.)}
\end{figure*}

We can now effect a transformation, denoted by $\Gd{\ddelta_0}'$, on the entire chain by displacing the polymer end $\rr_0$ by $\ddelta_0$ so that $\rr'_0 = \rr_0 + \ddelta_0$, then successively applying the discrete tractrix transformation with constant $f$ to every bond $\ttt_1$, $\ttt_2$, $\ldots$, $\ttt_N$ in that order, each time setting $\ddelta_{i - 1} = \rr'_{i - 1} - \rr_{i - 1}$.  Eventually, we obtain a new conformation of the chain: $\left\{\rr'_0,\,\rr'_1,\,\ldots,\,\rr'_N\right\} = \Gd{\ddelta_0}'\pullback{3}\left\{\rr_0,\,\rr_1,\,\ldots,\,\rr_N\right\}$.  Moreover we find that $\Gd{\ddelta_0}'$ is reversible since to recover the original configuration all we need to do is apply $\Gd{\textrm{--}\ddelta_0}'$ to the new configuration.  Another important feature of this transformation is that in general the end-to-end vector $\RR = \rr_0 - \rr_N$ changes, albeit not by $\ddelta_0$, since the other end $\rr_N$ is also displaced in the process.  However, suppose we wanted to change the end-to-end vector of the polymer by exactly $\DDelta$, there may exist some $\ddelta_0 = \ddelta^*$ for which this is possible, i.e., $\rr_0 + \DDelta - \rr_N = \rr_0 + \ddelta^* - \rr'_N$.

Thus we may address the aforementioned problem of the anchored polymer chain by temporarily setting it free, then determining $\Gd{\ddelta^*}'$ for which $\RR$ is incremented by exactly $\DDelta$.  After applying $\Gd{\ddelta^*}'$, we {\it rigidly} translate the entire chain by a displacement $\rr_N - \rr'_N$ so that the tail end which was temporarily set free coincides once more with its previous position.  We summarize the full transformation $\{\rr_0^{\left(s\right)},\,\rr_1^{\left(s\right)},\,\ldots,\,\rr_{N-1}^{\left(s\right)}\} \stackrel{\Gd{\DDelta}}{\longrightarrow} \{ \rr_0^{\left(s + 1\right)},\,\rr_1^{\left(s + 1\right)},\,\ldots,\,\rr_{N-1}^{\left(s + 1\right)} \}$ of the anchored linear discretized chain as follows:
\begin{eqnarray}
\label{eqn:transform}  \rr^{(s + 1)}_{0} &=& \rr^{(s)}_0 + \ddelta^* + (\rr^{(s)}_N - {\rr'}^{(s)}_N )
\,=\, \rr^{(s)}_0 + \DDelta, \\
\label{eqn:transform1}\rr^{(s + 1)}_{i} &=& {\rr'}^{(s)}_{i}  + (\rr^{(s)}_N - {\rr'}^{(s)}_N ), \quad  0 < i < N,
\end{eqnarray}
where $\ddelta^*$ solves the second equality in Eq.~\eqref{eqn:transform}, and
\begin{eqnarray}
\label{eqn:transformB}  {\rr'}^{(s)}_{0} &=& \rr^{(s)}_0 + \ddelta^*, \nonumber\\
{\rr'}^{(s)}_{i} &=& \mbg_{{}_{\textrm{T}}}\pullback{3}(\rr^{(s)}_{i},\, {\rr'}^{(s)}_{i - 1},\, {\rr'}^{(s)}_{i - 1} - \rr^{(s)}_{i - 1},\, f ),\;\;  0 < i \leq N.
\end{eqnarray}
In $n$ dimensions, Eq.~\eqref{eqn:transform} is a system of $n$ nonlinear equations in $n$ unknowns which may be solved for $\ddelta^*$.
The acceptance ratio of $\Gd{\DDelta}$ is given by
\begin{eqnarray}\label{acceptanceR}
\alpha_{\left(s\right) \rightarrow \left(s + 1\right)} &=& \frac{\textrm{e}^{-\beta \energy(\rr_k^{\left(s + 1\right)})}}{\textrm{e}^{-\beta \energy(\rr_k^{\left(s\right)})}} \left| \frac{\partial({\rr}_1^{\left(s + 1\right)},\,\ldots,\,{\rr}_{N - 1}^{\left(s + 1\right)})}{\partial(\rr_1^{\left(s\right)},\,\ldots,\,\rr_{N - 1}^{\left(s\right)})}\right|,
\end{eqnarray}
the last factor being the Jacobian determinant $\det\pullback{2}\left(\JJ_{\Gd{\DDelta}}\right)$ of the transformation.  Notice that we have used the fact that $\frac{\partial{\rr}_0^{\left(s + 1\right)}}{\partial\rr_k^{\left(s\right)}} = \delta_{0k}\IDM$, where $\IDM$ is the $n \times n$ identity matrix and $\delta_{ij}$  is the kronecker-delta [see Eq.~\eqref{eqn:transform}], to eliminate the first $n$ rows and columns of $\JJ_{\Gd{\DDelta}}$ as they do not contribute to the value of the determinant.  The key to computing $\JJ_{\Gd{\DDelta}}$ lies in first differentiating Eq.~\eqref{eqn:transform} with respect to $\rr_k^{\left(s\right)}$ and solving it to obtain:
\begin{eqnarray}
\label{forJac} \left.\frac{\partial{\rr'}_0^{\left(s \right)}}{\partial\rr_k^{\left(s\right)}}\right|_{\DDelta} &=& \left[ \IDM -  \left.\frac{\partial{\rr'}_N^{\left(s \right)}}{\partial{\rr'}^{(s)}_0}\right|_{\pullback{2}\{\pullback{2}\rr_j^{(s)}\pullback{2}\}}\pullback{3}\right]^{\textrm{--}1} \pullback{10}\cdot \left.\frac{\partial{\rr'}_N^{\left(s \right)}}{\partial\rr_k^{\left(s\right)}}\right|_{\ddelta_0 = \delta*} \pullback{23}+ \left(\delta_{0k} - \delta_{Nk} \right) \IDM \nonumber
\end{eqnarray}
where both derivatives in the right-hand-side may be found by differentiating Eq.~\eqref{eqn:BondTransform} appropriately.  The other derivatives $\left.\frac{\partial{\rr'}_i^{\left(s \right)}}{\partial\rr_k^{\left(s\right)}}\right|_{\DDelta}$ for $i = 1, \ldots N$ follow recursively from the latter equation after differentiating Eq.~\eqref{eqn:transformB}.  Finally, one may obtain all the matrix elements of $J_{\Gd{\DDelta}}$ by differentiating Eq.~\eqref{eqn:transform1} and substituting these results.

The determinant itself may be computed numerically by using $LU$-decomposition which has a complexity of $O(N^3)$.  The algorithm is fully scalable --- a suitable cut-off for the number of bonds $N$ taking part in the discrete tractrix move may be chosen beforehand to suit the speed of the computer.

Typically, for one simulation step, a central node is picked at random and displaced by $\DDelta$.  The corresponding $\ddelta^*$ for each arm originating from the central node is numerically solved to a specified precision and $\Gd{\DDelta}$ applied.  If no solution for $\ddelta^*$ is found for an arm, or if the numerical solver cannot reproduce the original configuration after the inverse transformation $\Gd{-\DDelta}$ is applied, then the entire simulation step is rejected in accordance with the Metropolis algorithm rules \cite{frenkel2001}, otherwise the complete acceptance ratio for the star's deformation is found by computing its total change in energy, the product of its arms' Jacobian determinants, and finally plugging them into Eq.~\eqref{acceptanceR}.  To ensure ergodicity, a few steps with random crank-shaft rotations can be applied to each arm between tractrix moves --- the crank-shaft rotations will enable each linking linear chain between nodes of the network to explore more of its possible conformations as the nodes remain fixed in space, while the discrete tractrix moves will displace the nodes themselves. We will outline several efficient variations of TRACTRIX in a future publication.

Our algorithm is the central result of this paper. To test it, we have applied it to various freely-jointed architectures ($\varepsilon = 0$) with both large and small numbers of bonds. In each case, TRACTRIX reproduced the equilibrium end-to-end distance distributions in exact agreement with their predicted analytical results. For systems where no such results exist, we conclude with a simple demonstration of TRACTRIX' proper functioning and use: Fig.~\ref{figPDF:wide} shows the results of simulations of various cross-linked WLC architectures, demonstrating an effect of some biological significance: cross-linked supramolecular polymers show a rising {\em effective} persistence length $\lp^{\pullback{4}*}$ with cross-linking density $n_\times$.  
Here $\lp^{\pullback{4}*}$ is defined as the persistence length of that WLC which has the same $\lc$ and expected value of the end-to-end distance $\avg{r} \equiv [\int \dx{r}\,r P(r)/(4\pi r^2)]/[\int \dx{r}\, P(r)/(4\pi r^2)]$ as the $n_\times$-cross-linked bundle [$P(r)$ denotes the end-to-end distance distribution].  Using this design motif, Nature may create supramolecular filaments of tunable effective stiffness with only two kinds of molecules at its disposal: identical chains and cross-linkers applied in varying concentrations.  Note, too, that $P(r)$ of an $n_\times$-cross-linked bundle is {\em not} the same as that of a single WLC with $\lp=\lp^{\pullback{4}*}(n_\times)$ --- in fact, one may have to use an extensible WLC variant to capture the complete effective mechanics. Though the stiffness of a bundle in reality also depends on the cross-linkers' stiffness and size \cite{Heussinger2007}, we did not consider this dependence in this initial survey.  Nor did we consider the effect of excluded volume interactions between chains, which would result in further stiffening each bundle while incurring the additional computational cost of having to reject all MCMC moves that cause filaments of now finite cross-section to overlap or violate topological constraints.  Extensive simulations of the collagen fibril, a supramolecular assembly of polypeptide triple helices are underway, and will be reported on in an upcoming publication. We emphasize that, although this work was inspired by bio-polymeric structures, TRACTRIX is in fact capable of dealing with similar configuration-space constraints in much more general settings and as such may find use well beyond biological polymers.

{\em Acknowledgements.} It is a pleasure to thank Prof. Daan Frenkel and Prof. Gerard Barkema for helpful discussions. This work is part of the Industrial Partnership Programme (IPP) Bio(-Related) Materials (BRM) of the Stichting voor Fundamenteel Onderzoek der Materie (FOM), which is supported financially by Nederlandse Organisatie voor Wetenschappelijk Onderzoek (NWO). The IPP BRM is co-financed by the Top Institute Food and Nutrition and the Dutch Polymer Institute. CS also acknowledges the hospitality of the Aspen Center for Physics where part of this work was conceived.


\begin{thebibliography}{99}
\bibitem{note1}
Coincidentally, this problem is equivalent to the so-called ``inverse kinematics problem'' which arises in robotics.
\end{thebibliography}
\end{document}